\documentclass{article}

\usepackage{arxiv}

\usepackage[utf8]{inputenc} 
\usepackage[T1]{fontenc}    
\usepackage{hyperref}       
\usepackage{url}            
\usepackage{booktabs}       
\usepackage{amsfonts}       
\usepackage{nicefrac}       
\usepackage{microtype}      
\usepackage{lipsum}
\usepackage{graphicx}
\graphicspath{ {./images/} }

\title{Ultrasensitive Doppler Raman spectroscopy using radio frequency phase shift detection}

\author{
 David R. Smith \\
  School of Biomedical Engineering\\
  Colorado State University\\
  Fort Collins, CO, 80523, USA\\
  \texttt{smithd@engr.colostate.edu} \\
   \And
 Jeffrey J. Field \\
  Department of Electrical and Computer Engineering\\
  Colorado State University\\
  Fort Collins, CO, 80523, USA\\
  \And
 David G. Winters \\
  Department of Electrical and Computer Engineering\\
  Colorado State University\\
  Fort Collins, CO, 80523, USA \\
  \And
 Scott Domingue \\
  Department of Electrical and Computer Engineering\\
  Colorado State University\\
  Fort Collins, CO, 80523, USA \\
  \And
 Frauke Rininsland \\
  Mesa Photonics, Inc.\\
  Santa Fe, NM, 87505, USA \\
  \And
Daniel J. Kane \\
  Mesa Photonics, Inc.\\
  Santa Fe, NM, 87505, USA \\
  \And  
 Jesse W. Wilson \\
  Department of Electrical and Computer Engineering\\
  Colorado State University\\
  Fort Collins, CO, 80523, USA \\
  \And
 Randy A. Bartels \\
  Department of Electrical and Computer Engineering\\
  Colorado State University\\
  Fort Collins, CO, 80523, USA \\
  \texttt{Randy.Bartels@ColoState.edu}\\
}

\begin{document}
\maketitle
\begin{abstract}
We introduce the first method to enable an optical amplification of a coherent Raman spectroscopy signal called radio frequency Doppler Raman spectroscopy. Doppler Raman measurements amplify the optical signals in coherent Raman spectroscopy by converting a spectral frequency shift imparted by an impulsive coherent Raman excitation to a change in a probe pulse transit time. This transit time perturbation is detected through the phase of a radio frequency electronic signal measured at a harmonic of the probe pulse train. By exploiting this new capability to scale the signal of a coherent Raman spectroscopic signal, we open the potential to detect very weak Raman spectroscopy signals that are currently not observable due to limits of illumination intensity imposed by laser damage to the specimen. 
\end{abstract}


\section{Introduction}
Optical microscopy is a key tool in many applications and is particularly valuable in biology where the fate and interaction of specific biomolecules must be tracked to understand processes that control the behavior of cells, tissues, and organisms. Fluorescent microscopic imaging using fluorescent probes tagged to particular molecules enables an astonishing range of biomedical studies \cite{Thorn:2016ef}. However, in many instances, fluorescent probes are too large to tag molecules and will modify diffusion properties and binding interactions with other molecules. Fluorescent probes face other issues such as photobleaching and technical difficulties in delivering fluorescent labels to target molecules. Challenges faced by labeling and fluorescent molecule stability have stimulated intense development of label-free molecular imaging techniques \cite{Ounkomol:2018aa}.

Other light spectroscopic interactions have long been pursued for imaging the behavior of biological molecules. Of particular interest is Raman spectroscopy, which can be used to differentiate molecules based on the intramolecular vibrational frequencies in the spectrum of inelastic light scattering from molecules in the specimen. Raman scattering is generally safe for biomedical imaging as the visible or near infrared light used for measurements is non-ionizing, though low-level multiphoton ionization imposes an upper limit on optical peak intensity\cite{Baumgart2009} and therefore limits sensitivity to low-concentration analytes. Use of visible and NIR light also enables excellent imaging resolution with Raman microscopy due to the short light wavelength and the abundance of high quality microscope objectives. 

\begin{figure}[ht]
\begin{center}
\resizebox{0.95\linewidth}{!}{\includegraphics{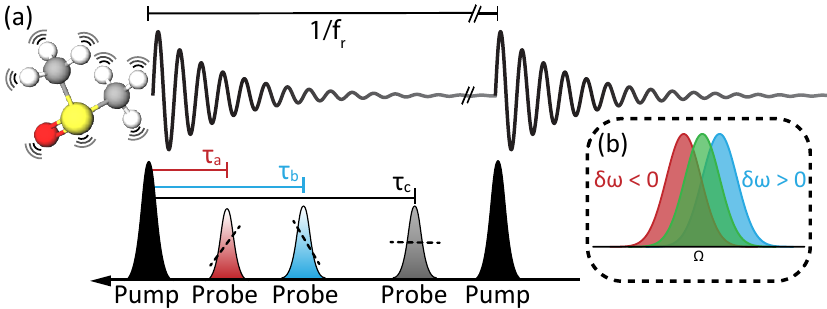}}
\caption{\label{fig:deltan} (a) A short laser pump pulse excited coherent vibrational motion leading to a time-varying perturbation of the optical susceptibility $\delta \chi^{(1)}(t;\tau) \approx 2 n_{\rm pr} \delta n(t;\tau)$, where $n_{\rm pr}$ is the refractive index experienced by the probe pulse at equilibrium. A time-delayed probe pulse arriving at a delay $\tau$ after the pump pulse experiences a time-varying linear phase modulation of $\phi_{\rm mod}(t;\tau) \approx k_{\rm pr} \, \delta n(t;\tau) \ell_f$. (b) The centroid of the power spectrum of the probe pulse train is shifted by an amount given by the local slope of the accumulated phase modulation $\delta \omega(\tau) \propto- \partial \phi_{\rm mod}/\partial t$ at the pump-probe pulse delay $\tau$.}
\end{center}
\end{figure}

Conventional Raman microscopy based on spontaneous Raman scattering has been widely used for cellular, tissue, in vivo, and plant imaging \cite{Gierlinger:2012wq, Chan:2008a}. However, spontaneous Raman faces severe limitations. Light interacts weakly with Raman vibrations, which is represented by the small Raman cross section per molecule -- on the order of $\sigma_R \approx 10^{-30} \,  {\rm cm}^2$. This weak cross section translates into low scattering rates. For example, illumination of a molecule with a 1 Watt light beam focused to a diameter of $150 \, \mu$m leads to less than one photon scattered by the spontaneous Raman process every hour. 

A Raman microscope with a high numerical aperture (NA) objective produces an intense focus that increases Raman scattering rates. Since the focused intensity must be kept below the damage threshold, integration times of seconds to minutes are necessary in order to acquire Raman microscope images. Raman spectra are usually contaminated by fluorescent light emission, which further limits the ability to extract signals above the noise level in Raman microscopy. These challenges prevent the observation of many molecules with spontaneous Raman techniques. Fluorescent background interference is particularly problematic when studying plant-derived specimens \cite{Gierlinger:2012wq}.

The limitations of spontaneous Raman scattering motivated the development of coherent Raman scattering (CRS) microscopy \cite{Duncan1982}. CRS provides a substantial increase in the rate of the scattering of incident light from the Raman interaction by coherently driving vibrational motion in a stimulated Raman process. By sending in pump and Stokes light fields that are separated by the frequency of a Raman vibration, the beat frequency of the fields drives strong vibrations in the molecule, producing an amplitude significantly larger than thermally excited vibrations. CRS scattering rates are orders of magnitude higher than spontaneous Raman scattering rates.

Efficiently driving CRS processes often requires phase matching, which limited early efforts at CRS microscopy \cite{Duncan1982}. After the demonstration that phase matching automatically occurs in the focus of a high NA objective \cite{Zumbusch1999}, CRS microscope technology has flourished \cite{ChenScience2015}. A set of powerful label-free imaging methods with excellent chemical specificity have emerged. Large pump and Stokes field intensities boost the CRS signal to noise ratio (SNR) in the recorded image \cite{ChenScience2015}. This SNR enhancement permits low pixel integration times -- thereby enabling high speed dynamic chemical imaging that has opened new applications.

However, the weak intrinsic scattering rate of Raman vibrations still prevents Raman spectroscopy and imaging in low concentration scenarios. While CRS signal strength can be increased with high intensity pump and Stokes fields, damage from the intense lasers prevents further scaling of the intensity \cite{Fu:2006zm}. Biological applications of CRS have mostly been limited to lipids, proteins (amide band), and DNA because these are the materials with the highest concentrations in cells.  Molecules with low Raman scattering cross sections and concentrations fall below the limit of detection for CRS microscopy and cannot be observed.  Many important biologically relevant molecules fall into this category, such as cytochromes, metabolites, and neurotransmitters. 

Resonant Raman scattering is able to probe Raman spectra of weak and low concentration molecules, but employs UV light that can be toxic to cells, and the SNR is still limited by the fluorescent background. Large enhancements from surface enhanced Raman scattering (SERS) based on field enhancements near the surface of structured metals enables single molecule detection of Raman scattering \cite{Stiles2008}. Unfortunately, applications of SERS methods are quite limited because analyte molecules must be in close proximity to the metal surface. 

In this Article, we demonstrate a new technique for measuring stimulated Raman spectroscopy whereby the amplitude of a coherently stimulated Raman signal is increased by an independent optical system. Our approach is based on impulsive stimulated Raman excitation (ISRE), in which a short pump pulse, with a duration $\tau_{\rm pu}$, efficiently excites coherent molecular vibrations in Raman-active modes with vibrational periods, $T_v = 2 \pi /\Omega_v$, shorter than the pulse duration \cite{Yan:1985uk}.  After the arrival of the pump pulse at $\tau=0$, the impulsively excited coherent molecular vibrations produce a time-varying perturbation to the linear optical susceptibility,  $\delta \chi^{(1)}(t;\tau)$. The Raman vibrational coherence can be probed by measuring the transient change in optical susceptibility by a time-delayed probe pulse that is detected through Kerr lensing \cite{Raanan:2018df}, diffraction \cite{Yan:1985uk}, an accumulated phase shift  \cite{Bartels:2002le, Wahlstrand:05, Wilson:2008lk, Hartinger:2008oe, Wilson:2008hh, Schlup:2009bq}, or a shift in the center frequency of the probe pulse spectrum \cite{Chesnoy:1988a, Domingue:2014bx}, i.e., a coherent Raman response-induced Doppler shift of the probe pulse spectrum, as illustrated in Fig. (\ref{fig:deltan}); we refer to this approach as Doppler Raman (DR).

\section{Radio frequency Doppler Raman (RFDR) spectroscopy}
Here we demonstrate the first method that can increase the signal strength of a Raman frequency shift outside of the interaction region, where laser damage currently limits the signal strength. To reach improved Raman detection sensitivity, we leverage advances in the measurement of timing jitter from precision optical metrology \cite{Scott:2001rb, Kim:2010hs} to detect small ISRE-induced Doppler frequency shifts \cite{DRPatent}. In order to apply the precision metrology methods, we convert the optical Doppler frequency shift imparted to the probe pulse train, $\delta \omega$, into a change in transit time $\Delta \tau$ using a dispersive element (Fig. 2d). By adapting methods employed for precision timing jitter metrology to measuring the DR shift induced change in transit time, very sensitive Raman spectroscopy measurements are possible.

Extremely low concentration detection through Raman spectroscopy is feasible because mode-locked laser oscillators exhibit extraordinarily low levels of timing jitter \cite{Jung:15}. Timing jitter measurements of mode-locked laser pulse trains with timing jitter $< 1$ as ($10^{-18}$ s) have been demonstrated \cite{Scott:2001rb}. This low jitter background noise implies that our DR spectroscopy technique will be able to detect frequency shifts of $\delta \nu = \delta \omega/2 \pi < 1$ kHz -- enabling the detection of molecules in low concentration and with weak Raman scattering cross sections.

A quantitative understanding of DR spectroscopy  requires a model for the excited time-varying optical susceptibility and subsequent probe pulse interaction.  For a single vibrational mode with frequency $\Omega_v$, the time-varying optical susceptibility takes the form $\delta\chi^{(1)}(t;\tau) = \delta \chi^{(1)}_0 F_v(t-\tau)$, where $\delta \chi^{(1)}_0$ is the peak change in susceptibility , $F_v(t) = \Theta(t) \exp(- \Gamma_v t) \sin(\Omega_v t + \phi_v)$, $\Gamma_v$ is the linewidth of the vibrational resonance, $\phi_v$ is a phase shift imposed by ISR excitation, and $\Theta(t)$ is the Heaviside step function that enforces causality. The expression for the optical susceptibility perturbation assumes a laser pulse with a peak pump pulse intensity,  $I_{\rm 0,pu}$, and a pulse duration, $\tau_{\rm pu}$, shorter than the excited vibrational modes. 

Raman spectra are acquired following ISRE by scanning the arrival time, $\tau$, of a probe pulse after impulsive excitation by the pump pulse. The probe pulse propagating at a delay time $\tau$ accumulates a phase shift given by $\delta \phi (\tau) = \delta \phi_0 F_v(\tau)$, where the peak phase shift, $\delta \phi_0 = k_{\rm pr} ~ \ell_f ~ \delta n(0;\tau)$, imparted onto the probe pulse for the Gaussian pump-pulse model reads \cite{Wilson:2012wc}
\begin{equation}
\delta \phi_0 = - \frac{2}{\tilde{\Omega}_v f_R} \, \beta_{\rm ISRE} \, g \, {\rm Im}\{ \chi^{(3)} (\Omega_v) \} \, p_{\rm pu}
\label{eq:PeakPhase}
\end{equation}
with
\begin{equation}
g = \frac{6 \pi}{n_{\rm pu} \, n_{\rm pr }  \,  A_f  \, c  \, \epsilon_0} \left( \frac{\ell_f}{\lambda_{\rm pr}} \right)
\label{eq:g}
\end{equation}
Here $\tilde{\Omega}_v = \sqrt{\Omega_v^2-\Gamma_v^2}$ is the reduced vibrational frequency, $\chi^{(3)} (\Omega_v)$ is the third order nonlinear susceptibility evaluated at the resonance frequency of the molecular vibration, $\beta_{\rm ISRE} = \tilde{\Omega}_v \Gamma_v \exp[-(\Omega_v \tau_{\rm pu}/2)^2]$ is the ISRE parameter, and $f_R$ is the repetition rate of the pump and probe pulse trains. We have denoted the pump pulse train average power, interaction length of the focused probe beam, cross sectional focal beam area, and pump pulse refractive index as $p_{\rm pu}$, $\ell_f$, $A_f$, and $n_{\rm pu}$, respectively, and $n_{\rm pr}$ and $\lambda_{\rm pr}$ are the refractive index and the center wavelength of the incident probe pulse. 

The time-varying phase shift accumulated by the probe pulse, $\delta \phi_{\rm mod}(t)$, is proportional to a temporally modulated optical path length (OPL). A temporal variation in the OPL imparts a Doppler frequency shift to the probe pulse given by $\delta \omega = - k_{\rm pr} \, \partial \, {\rm OPL}/\partial \, t$, where $k_{\rm pr} = 2 \, \pi/\lambda_{\rm pr}$ is the wavenumber of the probe pulse, with $\lambda_{\rm pr}$ denoting the wavelength of the probe pulse center frequency. As ${\rm OPL} = n \, \ell$  is the product of the refractive index and the physical propagation distance, it follows from the chain rule that there are two possible origins of a Doppler shift. The common case of a moving scattering particle leads to $\delta \omega = k_{\rm pr} \, n \, \partial \, \ell/\partial \, t$, where the Doppler shift originates from a time-varying physical path length scattered from a moving object and produces a large change in the fringe density of an interferometric measurement, and is the reason that Doppler OCT can readily measure small frequency shifts imparted on an ultrafast laser pulse train \cite{Chen:1997dz}.  However, when a phase modulation is accumulated by a rapid change in refractive index, producing small frequency shifts $\delta \omega = k_{\rm pr} \, \ell \, \partial \, n/\partial \, t$, such frequency shifts produce a signal change too weak to be measured in an interferometer. This generalized Doppler shift process was first observed due to frequency shifts of satellite microwave pulses propagating through the turbulent atmosphere \cite{Hopfield:2015ft}. In the case of DR, this generalized Doppler shift is uniform for every pulse in the probe pulse train because the molecular system relaxes back to thermal equilibrium in between each pump pulse, so that an identical frequency shift is imparted on every pulse in the laser pulse train for equivalent ISRE conditions.
 
ISRE spectra have been recorded by measuring the probe pulse diffraction from a transient grating formed by a pair of pump pulses \cite{Yan:1985uk} or Kerr lensing \cite{Raanan:2018df} as a function of pump-probe delay time, $\tau$.  Raman spectra obtained by recording the ISRE-induced probe pulse phase shift have been obtained with both spatial \cite{Bartels:2002le} and spectral interferometry methods \cite{Wilson:2008hh, Wilson:2008lk, Schlup:2009bq}. In those interferometric measurements, it is assumed that the probe pulse duration $\tau_{pr}$ is shorter than the vibrational period $T_v$. When the probe pulse duration is $\tau_{pr}$ long compared to the vibrational period, the periodic phase modulation accumulated by the pulse produces equal-amplitude Stokes and anti-Stokes sidebands on the probe pulse spectrum \cite{Weinacht:2001oj, Bartels:2002le}.

\begin{figure*}[ht]
\begin{center}
\resizebox{0.95\linewidth}{!}{\includegraphics{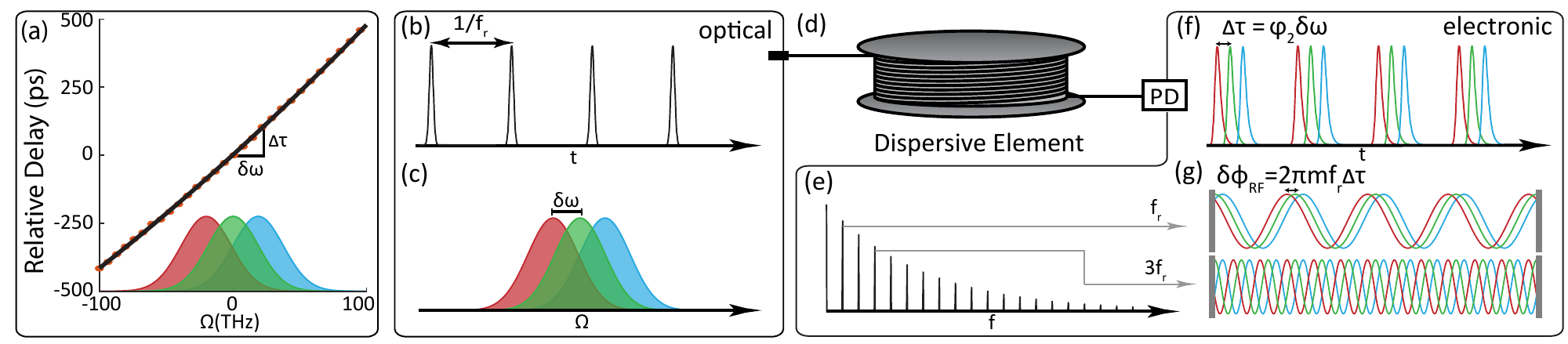}}
\caption{\label{fig:frequencytotime} (a) A dispersive optical system maps the frequency shift of the probe pulse to a change in transit time through the optical fiber due to group delay dispersion ${\rm GDD} = \varphi_2$. The probe pulse optical pulse train (b) accumulates a frequency shift (c) due to propagation through the vibrational coherence prepared by the pump pulse. (d) The frequency shift is converted to a change in time of flight through a length of multimode fiber (MMF) whereupon a photodiode (f) converts the optical pulse train to an electronic pulse train. The m$^{\rm th}$ harmonic of the pulse repetition frequency (e) is isolated electronically and a phase detector is used to record the RF phase shift (g), which is the measurement of the Raman excitation.}
\end{center}
\end{figure*}

Near the center of the probe pulse at a delay $(t-\tau)$, there is a linear phase ramp that imparts a Doppler frequency shift on the centroid of the optical spectrum given by $\delta \omega \approx \delta \omega_0 \, G_v(\tau)$. The peak frequency shift is $\delta \omega_0 = \delta \phi_0 ~ \Omega_v$ and the delay dependence of $G_v(\tau) \approx \exp(- \Gamma_v \tau) \cos(\Omega_v \tau + \phi_v)$ is used to obtain the time-domain Raman spectral signal \cite{Bartels:2001wo, Gershgoren:2003tx}. The center frequency shift of the probe pulse can be converted into a change in transmitted pulse energy by passing the pulse through a spectral disperser. Many methods have been adopted to measure the Raman spectrum by recording pump-probe delay traces of the power transmitted through such a spectral filter \cite{Dhar:1994sy, Merlin:1997ca, Dudovich:2002cj, Domingue:2014bx, Raanan:2018wu} or through subsequent nonlinear signal generation \cite{Hartinger:2008qd, Kupka:2010qp}. The phase shifting, frequency shifting, and nonlinear frequency conversion methods are all time-domain Raman techniques, so that the complex Raman spectrum is obtained by Fourier transforming the recorded probe pulse signal. In particular, the method of recording the Raman spectrum with probe pulse transmission through a spectral filter is routinely used for measuring low-frequency optical phonon vibrational spectra \cite{Merlin:1997ca}. 

No efforts to date have pushed sensitivity limits of impulsive Raman with external optical amplification. In order to improve the sensitivity of low concentration molecule detection, with DR, we have developed a method of very sensitive detection of small frequency shifts imparted to an ultrafast laser pulse train. To motivate the need for small frequency shift detection, we note that the Doppler frequency shifts due to ISRE increase in direct proportion to the third-order Raman susceptibility, $\delta \omega_0 \propto {\rm Im}[\chi^{(3)}(\Omega_v)]$. The third-order optical susceptibility scales with the number density of molecules, $N$, as $\chi^{(3)}(\Omega_v) = N \, \langle \gamma^{(3)}(\Omega_v)\rangle/\epsilon_0$. Here $\epsilon_0$ is the dielectric permittivity of free vacuum and $\langle \gamma^{(3)}(\Omega_v) \rangle$ is the orientational averaged second hyperpolarizability  at the resonant vibrational frequency \cite{Hartinger:2008oe, Cleff:2016cz}. The second hyperpolarizability $\gamma^{(3)}(\Delta \Omega) = \gamma^{(3)}_{\rm NR} + \gamma^{(3)}_{\rm VR} \, f_v(\Delta \Omega)$ exhibits a non-resonant electronic contribution to the second hyperpolarizability $\gamma^{(3)}_{\rm NR}$ and a vibrational Raman response with an amplitude $\gamma^{(3)}_{\rm VR}(\Omega_v) = -(12 \, i \, \Gamma_v \, \Omega_v)/(\partial \alpha /\partial Q_v)_0$ at resonant vibrational frequency and a resonant spectral response of $f_v(\Delta \Omega) =  - 2 \, i \, \Gamma_v \, \Omega_v /(\Omega^2_v - \Delta \Omega^2 +  2 \, i \, \Gamma_v \Delta \Omega)$ \cite{Shen:1965hp}. 

The magnitude of the Raman response is characterized by the derived polarizability, $(\partial \alpha/\partial Q_v)_0$, of a vibrational mode. From the model, it is clear that the phase and frequency shifts are linearly proportional to the molecular concentration. A 25-fs pump pulse at a center wavelength of 800 nm and with 3 mW average power from a mode-locked laser with a 94 MHz repetition rate focused to a diffraction-limited spot that excited the 459 cm$^{-1}$ symmetric stretch mode in CCl$_4$ \cite{Kato:1971eb} will impart a maximum frequency shift of $\delta \omega_0/2 \, \pi \sim 130$ GHz on a probe pulse that is initially centered at 800 nm. This implies that if we were able to detect a frequency shift of less than 1 kHz, we would be able to probe molecular concentrations below 100 nM.  This is within the range of precision timing jitter metrology techniques.

In order to detect these DR frequency shifts with exceptional sensitivity, we developed a method to convert the Doppler frequency shift, $\delta \omega$, into a time delay, $\Delta \tau$, and then adapted a precision timing jitter metrology method to measure small changes in the arrival time of the probe pulse due to the interaction with the vibrational coherence in the specimen that was prepared by the pump pulse. The strategy for converting the frequency shift to a time shift is illustrated in Fig. (\ref{fig:frequencytotime}). We pass the probe pulse train that has accumulated a frequency shift of $\delta \omega$ through an optically dispersive system with a group delay $\tau_g$ that can be approximated as $\tau_g \approx \tau_{\rm g0} + \varphi_2 \, \delta \omega$, where $\varphi_2$ is the group delay dispersion (GDD) and $\tau_{\rm g0}$ is the transit time for probe pulses centered on the original center frequency $\omega_0$ through the dispersive optical system. A change in the center frequency of the probe pulse is thus converted into a change in transit time through the dispersive optical system, $\Delta \tau = \tau_g - \tau_{\rm g0} = \varphi_2 \, \delta \omega$. The GDD stretches the pulse in time to a chirped pulse duration $\tau_{\rm pr,c}$, and the GDD can be estimated as $\varphi_2 \approx \tau_{\rm pr,0} \, \tau_{\rm pr,c}$ for a strongly chirped pulse, $\tau_{\rm pr,c}$, and $\tau_{\rm pr,0}$ is the transform-limited probe pulse duration. The pulse duration should not exceed the pulse spacing of the pulse train, which sets a limit on the maximum GDD that can be imparted.

The induced DR timing jitter is modulated sinusoidally by modulating the pump pulse train energy with an acousto-optic modulator. The induced sinusoidal probe pulse timing jitter is detected with a standard method of timing jitter measurement through detection of the phase shift of a harmonic of the pulsed electronic signal that is produced in a photodiode \cite{McFerran2005}. The peak RF phase shift recorded for the $m^{\mathrm{th}}$ harmonic reads $\delta \phi_{\rm RFDR} = 2 \, \pi \, m \, f_R \, \Delta \tau$.

DR spectroscopy is able to detect low molecular concentrations due to the fact that mode-locked ultrafast lasers exhibit exceptionally low timing jitter (and thus phase noise at harmonics of the repetition rate of the laser) in the pulse train. Timing jitter power spectral densities (PSD) of $S_{\delta \tau} \sim 10^{-b}$ fs$^2/$Hz with $b>10$ at an offset frequency $> 1$ MHz have been measured  \cite{Jung:15}. This timing jitter PSD leads to an rms timing jitter of $\delta\tau_{\rm rms} <$ 1 as  in a 100 $\mu$s integration time. The minimum detectable frequency shift occurs when we reach a DR transit-time change equal to the timing jitter noise level, given by $\delta \omega_{\rm min} \sim \delta\tau_{\rm rms} / \varphi_2$. For this timing jitter PSD with sub-attosecond jitter, we expect a detectable frequency shift down to $\delta \nu < 1$ kHz, leading to an expected $<$ 100 nM concentration detection limit for CCl$_4$.

To appreciate the favorable limit of detection scaling for DR, we note that other sensitive methods of coherent Raman scattering use a pump-probe detection modality where a small power change of the probe pulse, $\Delta p$, is imparted by the Raman interaction. For Stimulated Raman Scattering (SRS) and for ISRE detection with a frequency filter, the fractional change in probe power reads $\Delta p/p_{\rm avg} \approx \delta\omega/\Delta \Omega_p$. The constant of proportionality is of order of unity. Depending on the relative intensity noise (RIN) of the laser and the average probe pulse power, the noise will be dominated by either Shot noise or RIN, and with low noise sources  $\Delta p/p_{\rm avg} \approx 10^{-6}$ can be detected. Thus, the equivalent frequency shift detection limit for a method that imparts a change in probe pulse power is on the order of $\delta \nu_{\rm min} \sim 10$ MHz. DR detection offers the potential for much more sensitive Raman detection due to the fact that the timing jitter of mode-locked lasers is significantly lower than the RIN.

In this Article, we present the first implementation of DR detection using direct electronic detection of the timing jitter through an RF phase shift. The detection of mode-locked laser timing jitter through purely electronic means introduces several noise sources that degrade the minimum detectable timing jitter value \cite{McFerran2005}. The lowest possible phase noise in an RF phase shift timing delay system is due to Shot noise and is inversely proportional to the RF electronic power, $p_{\rm RF}$, in the harmonic order used for phase shift detection in the photodiode. While additional noise sources are possible through electronic detection of timing fluctuations of a mode-locked laser pulse train, we operate at the RF phase noise detection Shot noise limit set by the photodetection process. However, proper operation makes RF phase detection largely insensitive to intensity noise. Below, we demonstrate Shot noise limited detection of RFDR spectroscopy.

\section{Results}

\begin{figure}[htbp]
	\begin{center}
		\resizebox{0.55\linewidth}{!}{\includegraphics[width=0.45\textwidth]{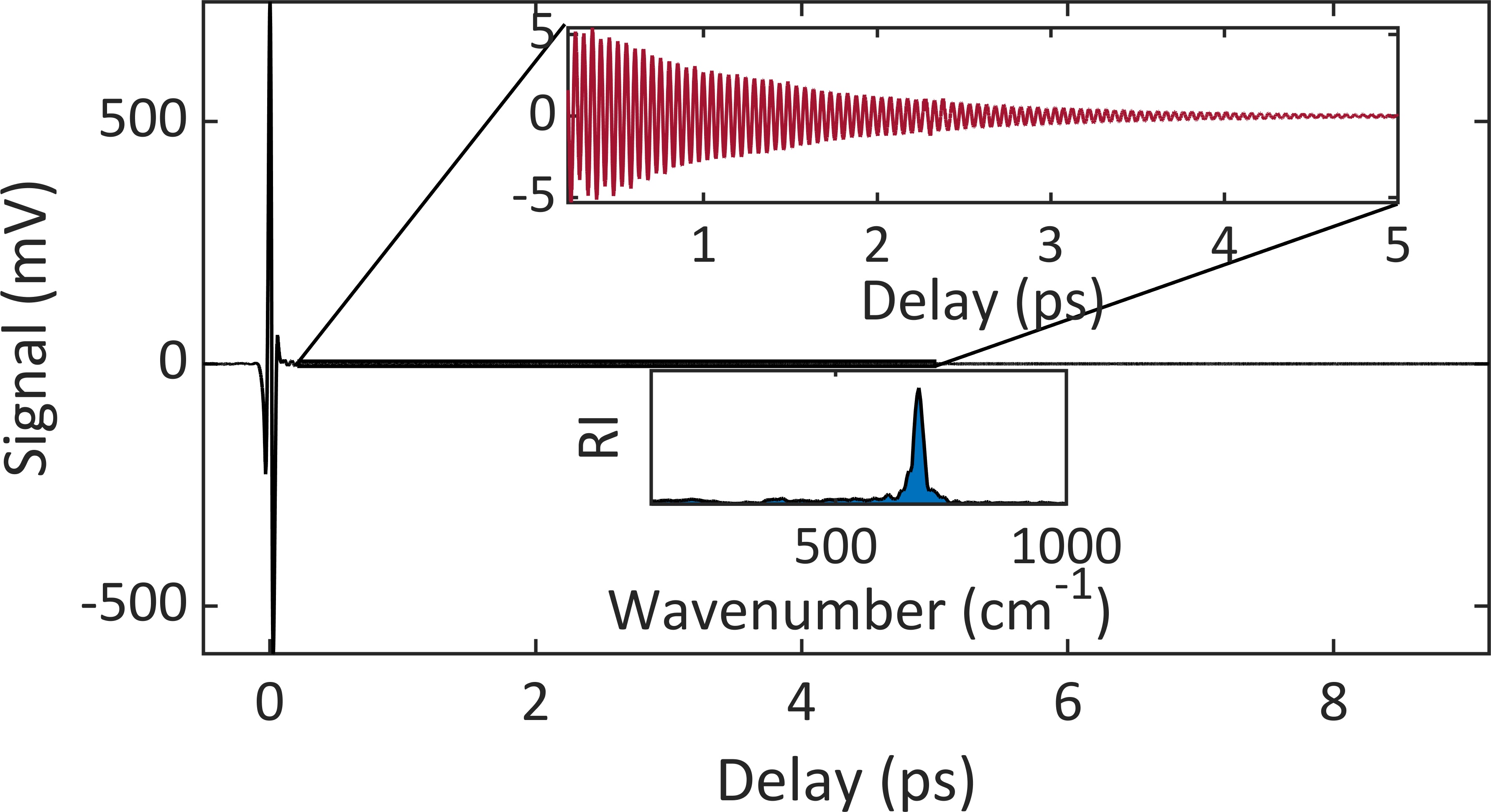}}
		\caption{\label{fig:dmso}  Time Resolved signal from 200mM Dimethyl Sulfoxide (DMSO).  The large cross phase modulation peak is visible with the Raman response (upper inset), highlighting the dynamic range of the measurement system.  The Raman spectrum recovered using multitaper power spectral density estimation is shown in the lower inset.}
	\end{center}
\end{figure}

\begin{figure}[ht]
\begin{center}
\resizebox{0.55\linewidth}{!}{\includegraphics[width=0.45\textwidth]{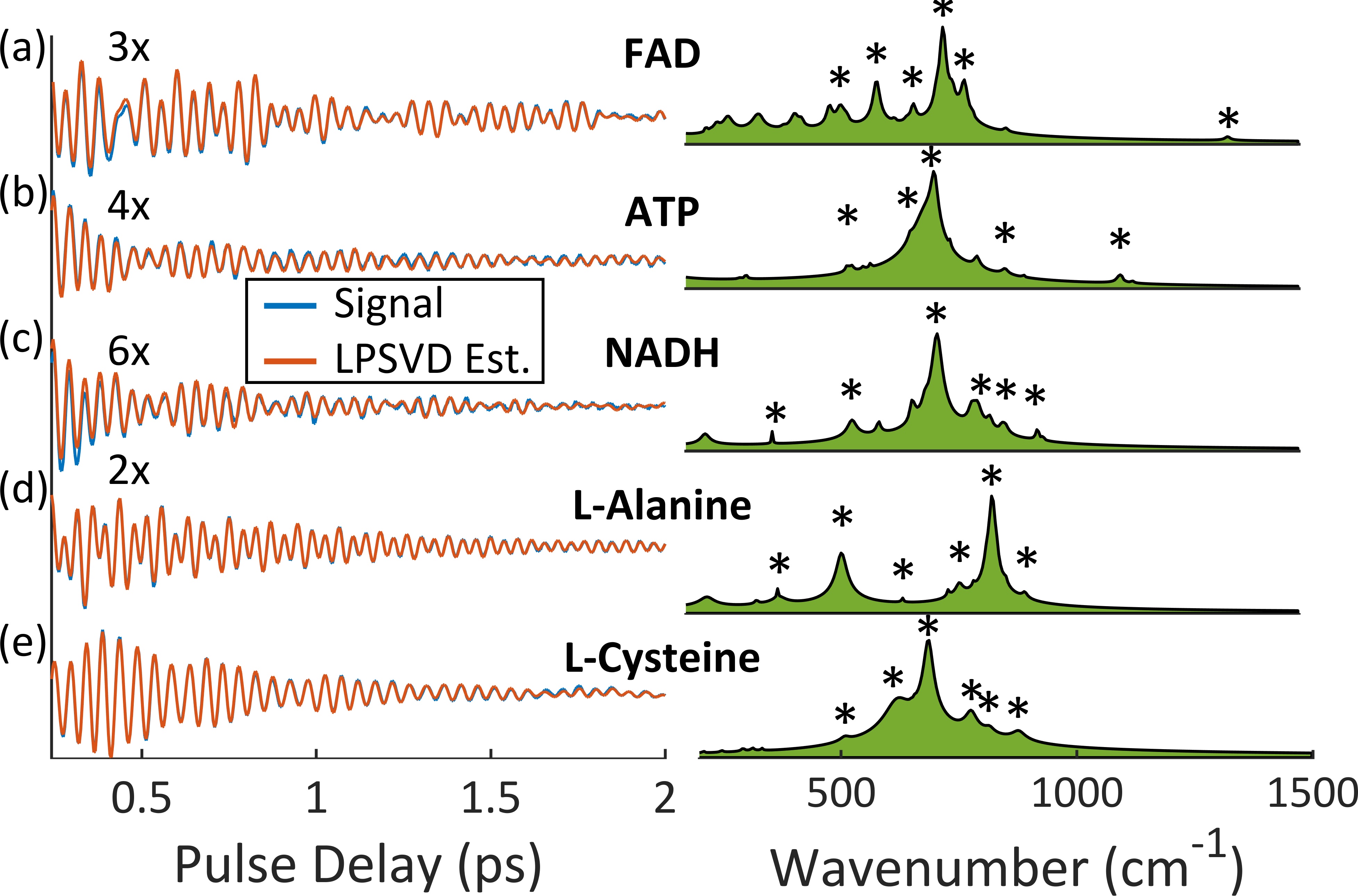}}
\caption{\label{fig:spectra} Time-resolved Raman spectra of biological molecules measured with RFDR spectroscopy. Signals with linear prediction singular value decomposition (LPSVD) model estimates are shown at left.  Raman spectra reconstructed using LPSVD are shown at right where peaks corresponding to previously reported literature values are marked. (a) 100mM Flavin adenine dinucleotide (FAD) in phosphate buffer saline (PBS). (b) 100mM Adenosine triphosphate in DI water.  (c) 100mM Nicotinamide adenine dinucleotide (NADH) in PBS. (d) 0.5M L-Alanine in PBS.  (e) 1M L-Cysteine in PBS.   }
\end{center}
\end{figure}

\begin{figure}[ht]
\begin{center}
\resizebox{0.55\linewidth}{!}{\includegraphics[width=0.45\textwidth]{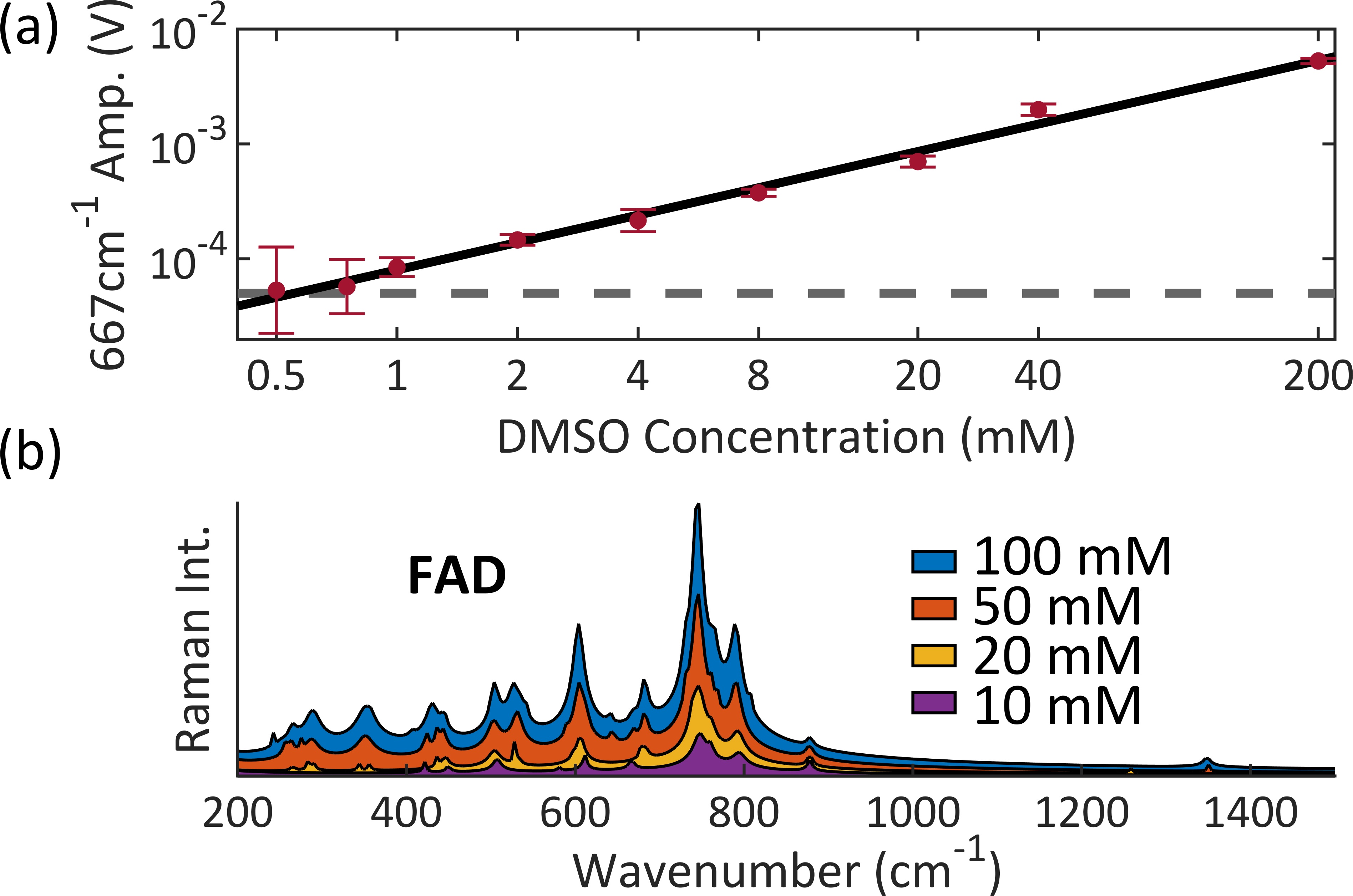}}
\caption{\label{fig:conc}  (a) Scaling of the RFDR signal strength of the 667~cm$^{-1}$ mode of dimethyl sulfoxide (DMSO) with concentration.  Dashed line represents the current system noise floor. (b) Flavin adenine dinucleotide (FAD) spectra are shown as a function of concentration. }
\end{center}
\end{figure}

\begin{figure}[ht]
	\begin{center}
		\resizebox{0.55\linewidth}{!}{\includegraphics[width=0.45\textwidth]{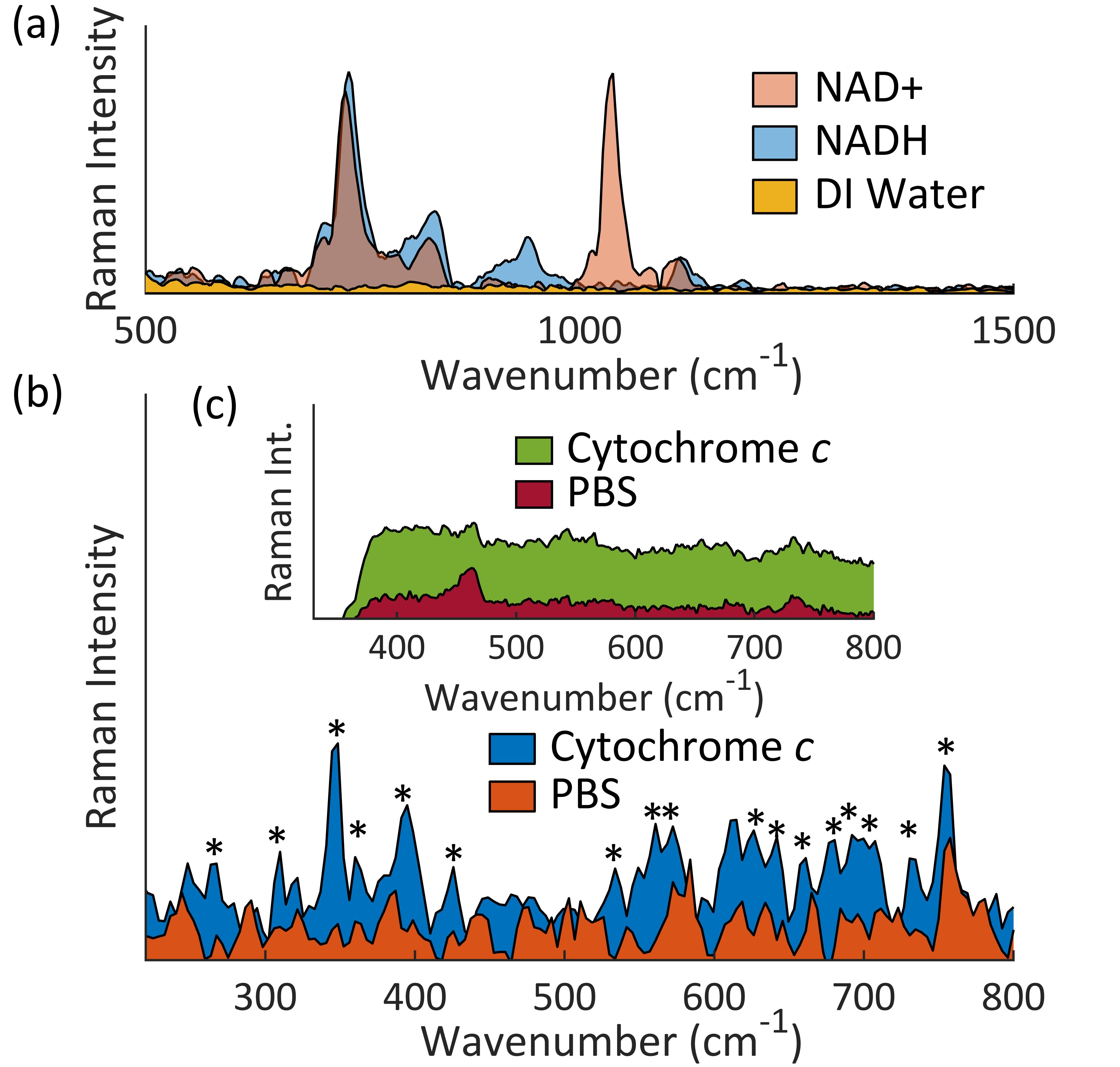}}
		\caption{\label{fig:nadh}  (a) DR spectra for NAD+ and its reduced form, NADH, are shown with a DI water background. (b) DR spectra from 2mM cytochrome \emph{c} in PBS. (c) Spontaneous Raman spectrum of 2mM cytochrome \emph{c} in PBS.}  
	\end{center}
\end{figure}

A detailed description of the experimental DR system is provided in the Methods section of the Supplemental Information. The amplitude of the DR RF phase shift as a function of pump-probe delay is stored digitally for analysis. Data reporting the amplitude of the modulation of the RF phase shift imparted by the modulated pump pulse energy that varied the excitation strength of the vibrational coherence are directly acquired from the lock-in amplifier as a function of pump-probe delay.  At a fixed pump-probe time delay, $\tau$, the time-varying strength of the excited vibrations caused a time-varying center frequency shift, and thus a time-varying change in the transit time of the probe pulse train through the optical fiber, which is finally reflected as a sinusoidal modulation of the RF harmonic phase. Thus, the amplitude of this swing in the RF phase shift imparted by the modulation of the pump pulse energy measures that same modulation in the vibrational coherence, and thus the molecular concentration. Recording this peak phase shift as a function of pump-probe delay enables the Raman spectrum of the vibrational modes to be recorded. Fig. (\ref{fig:dmso}) shows a typical time-resolved response for a Raman-active sample. Raman spectra are recovered from the time-resolved Raman response by estimating the signal power spectrum using multitaper power spectral density estimation or using linear prediction singular value decomposition (LPSVD); the power spectrum is shown in the inset of Fig. (\ref{fig:dmso}). The exceptional dynamic range of the DR system is demonstrated in the system's ability to resolve the large cross phase modulation signal at $\tau = 0$ while the upper inset shows the exponentially decaying Raman response of the sample and the lower inset shows the Raman spectrum of DMSO. The cross phase peak to the system noise floor represents a dynamic range of more than 5 orders of magnitude.  Raman spectra for a number of metabolites are shown in Fig. (\ref{fig:spectra}), and show excellent agreement with Raman spectral peaks from the literature, which are marked with an asterisk.  References for the spectral peaks can be found in the Supplementary Information.

The detected RF phase shift signal in DR is given by the simple formula, $\delta \phi_{\rm RF} = 2 \pi f_r m \varphi_2 \delta \omega$, that shows the signal is linearly proportional to the detected phase shift with harmonic order $m$ and the GDD ($\varphi_2$). Furthermore, the frequency shift, $\delta \omega \propto [C] $, is expected to scale linearly with the molecular concentration. We demonstrated the linear scaling of both of these quantities experimentally. In Supplemental Figure S1, the amplitude of the 90~cm$^{-1}$ mode of BGO was recorded for a wide range of values of GDD and harmonic order. For each harmonic order, the GDD is varied by changing the total length of the dispersive optical fiber, and the expected linear behavior is observed. These data also show the expected linear increase in signal with harmonic order, $m$. The RF phase measurements also depend on the parameters of the RF mixer used as a phase detector and details of the RF power levels used to saturate the mixers. Linearity of the scaling with harmonic order was verified by operating with a single mixer with carefully adjusted RF power levels to ensure similar mixer saturation conditions.

The linearity of the scaling of the DR spectroscopy signal with concentration is shown in Fig. (\ref{fig:conc} a). To establish the sensitivity limit of the current DR spectroscopy system, concentration studies with DMSO were performed.  DMSO was selected because it is easily diluted with DI water, which lacks a detectable Raman response with DR spectroscopy.  A series of dilutions were prepared using neat DMSO and DI water to obtain DMSO concentrations from 200 mM to 0.5 mM.  DR spectroscopy measurements were performed in a flow-through quartz cuvette that was mounted in the specimen plane of the spectroscope.  Using a syringe, 15 mL of a DMSO solution was flushed through the flow-through cuvette hosing lines, filling the cuvette with the solution.  Data was recorded before repeating the solution flushing three times for each concentration.  The syringe volume was sufficient to fully flush the previous solution from the flow-through cuvette before each data run.  15 mL of DI water was plunged through the cuvette between each concentration to prevent contamination from prior solutions.  Further, the concentration study started from the lowest concentration (0.5 mM) to further mitigate the potential of contamination from prior solutions. The pump and probe powers for the concentration study were 80 mW and 49 mW respectively. The time-resolved data was processed using LPSVD to recover the peak amplitude for the 667~cm$^{-1}$ mode of DMSO.  The rms noise floor shown with the dashed line in Fig. (\ref{fig:conc} a) was measured by performing a data run with DI water in the flow-through cuvette, and these values are consistent with independent measurements of the Shot noise limiting timing jitter noise shown in the supplemental information. Fig. (\ref{fig:conc} b) shows LPSVD-derived Raman spectra for a concentration study with FAD.

DR spectroscopy was used to observe differences in redox states of the biologically significant molecules nicotinamide adenine dinucleotide (NAD+).  The NAD+ and NADH solutions were prepared in DI water.  In Fig. (\ref{fig:nadh} a) significant differences in the spectra are seen, particularly the presence of the 1038~cm$^{-1}$ mode with NAD$^+$.  This result agrees well with previous studies \cite{Yue1986}.  Fig. (\ref{fig:nadh} b) shows DR spectra from 2mM cytochrome \emph{c} in PBS where peaks that agree with previous literature have been marked \cite{Hu1993}.  Fig. (\ref{fig:nadh} c) shows the spontaneous Raman spectrum from the same cytochrome \emph{c} sample.  The spontaneous Raman spectrum was recorded using a home-built spontaneous Raman microscope with pump wavelength $\lambda_{pump} = 532~\mathrm{nm}$. The spontaneous Raman spectrum is dominated by fluorescence background from the fluorescent oxidized form of cytochrome \emph{c}, masking any Raman peaks.  DR spectroscopy is immune to the fluorescence background that can be prohibitive with spontaneous Raman.  Similar to other ISRE-based systems, DR readily detects low frequency Raman modes that can be difficult to record with spontaneous Raman and other coherent Raman techniques.  The Raman spectra were recovered using multitaper power spectral density estimation.  

\section{Discussion}
The primary motivation of DR spectroscopy is to boost the signal-to-noise ratio (SNR) of Raman spectroscopy measurements to push to lower concentration detection limits and to detect molecules with very weak Raman cross sections. DR spectroscopic detection offers a unique pathway for ultrasensitive Raman detection because of two primary benefits. Firstly, the Raman signal is amplified in a dispersive medium when the small center frequency shift accumulated by a probe pulse is converted into a timing delay by the dispersion. By increasing the dispersion, the coherent Raman signal is optically amplified beyond the limits set by damage at the focus, where other coherent Raman techniques are constrained. This change in transit time is converted into a periodic timing jitter that is induced by modulating the pump pulse train power. By imparting a periodic DR timing jitter, this DR signal can be detected with high sensitivity using a lock-in amplifier at large offset frequencies that exhibit low noise. The second major advantage is that the noise floor for DR measurements is the timing jitter power spectral density, rather than the relative intensity noise (RIN) that sets the noise floor for all other CRS methods, such as stimulated Raman scattering (SRS) \cite{Zhang:2015bc} and spectral filtered impulsive stimulated Raman scattering. The conversion of the frequency shift to a time delay opens a new possibility for very low noise measurements because mode-locked ultrafast oscillators display exceptionally low timing jitter noise. 

The DR signal arises from the modulated time delay signal, $\Delta \tau$, whereas the timing jitter noise floor arises from the rms value of the timing jitter noise, $\delta \tau_{\rm rms}$, from the mode-locked laser oscillator, and ratio of these quantities defines the signal-to-noise ratio, ${\rm SNR} = \Delta \tau/\delta \tau_{\rm rms}$. The minimum detectable frequency shift, $\delta \nu_{\rm min}$, and thus the minimum detectable molecular concentration, are from setting the SNR to unity, giving  $\delta \nu_{\rm min} = \sqrt{S_{\delta \tau}(f_m)} / 2 \, \varphi_2 \, \sqrt{2 \Delta t}$. Here, we assume that the timing jitter noise PSD is approximately constant across the detector bandwidth $\Delta f = 1/2 \Delta t$, which is characterized by the integration time $\Delta t$ for the offset frequencies centered at the AOM modulation frequency imparted onto the pump pulse train, $f_m$. By comparison, other CRS methods record Raman spectra by detecting a change in optical probe power relative to the average probe pulse power. These are limited by the relative intensity noise (RIN) of the laser source. Measurements of the CRS spectrum through the power fluctuations lead to a minimum detected frequency shift of $\delta \nu_{\rm min}  =  10^{\sigma_{\rm RIN}/20} \,/ 2 \, \pi \, \tau_{\rm pr}$ when the probe pulse average power is large enough for RIN to exceed Shot noise.

In the DR spectroscopy system presented here, we use an electronic detection strategy where the RF phase of a harmonic of the repetition rate is detected to recover the ISRE frequency shift induced on the probe pulse. Our experiment operates in the limit where noise in the measurement is limited by Shot noise generated in the photodiode. The lowest measurable timing jitter PSD with electronics phase noise detection is set by the phase Shot noise set by the average RF power, $p_{\rm RF}$, in the m$^{th}$ harmonic of $f_R$, and is given by $S_{\delta \tau} = h \, \nu/ 2 \, p_{\rm RF} \, (\pi m f_R)^2 $. Here $h$ is Planck's constant and $\nu$ is the center optical frequency. Because the average probe power is constant and the average photocurrent is fixed, as the dispersion (parameterized by $\varphi_2$) of the frequency-to-time delay converter is increased to increase the DR signal, the probe pulse time duration also increases. The increased probe pulse durations with increasing $\varphi_2$ produces a nonlinear drop in RF power with increasing harmonic order $m$. As a result, the Shot noise, and commensurately $S_{\delta \tau}$, increases nonlinearly with increasing harmonic order.

The 500m segment of MMF used for most measurements in this work imparted a GDD of $\varphi_2 = 22.5$ ps$^2$ onto the probe pulses. This value of $\varphi_2$ produces a frequency shift to transit delay mapping of $\Delta \tau = 0.14$ fs/MHz. The minimum detectable frequency shift for these experimental conditions gives the expression $\delta \nu _{\rm \min} =\sqrt{S_{\delta \tau}}/ 2 \pi \sqrt{2} \varphi_2 \sqrt{\Delta t}$. The first and second harmonics performed similarly and most measurements were recoded at $m = 2$, the minimum detectable frequency shift for our experimental scenario is $\delta \nu _{\rm \min} =  1.2 ~ {\rm MHz} ~ {\rm s}^{0.5} / \sqrt{\Delta t}$. In our experiments, we find that averaging over 7 scans, we measure a minimum frequency shift of $\delta \nu _{\rm \min} = 9.08$ MHz, compared to the estimated minimum value of $8.54$ MHz determined by the Shot noise floor -- demonstrating operation of DR spectroscopy within 6$\%$ of the Shot noise limit for RF phase noise detection. 

To appreciate the advantage of the amplified time delay measurement of DR over conventional measurements that detect CRS signals through a change in probe power, we consider noise levels of various experimental systems. At the limiting value of GDD, where the probe pulse temporal duration reaches the pulse train separation $f^{-1}_R$, then $\varphi_2 = \tau_0 / f_R$, and we reach a maximum signal-to-noise value ${\rm SNR} =  \delta \omega \, \tau_0 \, \sqrt{2 \Delta t}/f_R \,\sqrt{S_{\delta \tau}(f_m)}$ for the detector integration time, $\Delta t$. Timing jitter PSD can be approximated as constant over the detector integration bandwidth, with a PSD of the form of $S_{\delta \tau} \approx 10^{-b}$ fs$^2$/Hz, where $b \sim 10$ can be a few MHz offset in mode-locked laser sources \cite{Jung:15}. Under typical conditions, the equivalent RIN required, $\sigma{\rm RIN} \approx -200$ dBc, is orders of magnitude lower than RIN levels of any laser source. Mode-locked lasers with high stability achieve $\sigma{\rm RIN} \approx -150$ dBc, which for an integration time of $\Delta t = 300 \mu$s, conventional CRS methods based on probe power pulse changes, such as SRS are and SFDR, are limited to a concentration limit of detection of approximately $[{\rm CCl}_4]_{\rm min} \approx 5$ mM. In contrast, RF DR accesses the low timing jitter noise floor, and for the same integration time, a concentration detection limit of $[{\rm CCl}_4]_{\rm min} \approx 24$ nM is feasible.

\section{Conclusions}
To conclude, we have demonstrated a new concept for coherent Raman spectroscopy that allows direct optical amplification of the Raman signal. In this work, the Raman signal is a center frequency shift of a time-delayed probe pulse that is converted into a transit time that scales with the GDD of the dispersive system. While the current measurements are limited by the Shot phase noise generated in a photodetector, this work demonstrates the potential of DR spectroscopy for unprecedented low concentration Raman detection for measurements by exploiting the exceptionally low timing jitter of mode-locked ultrafast lasers. Improved DR detection methods will open the possibility for few molecule DR detection without the need for local field enhancements. Moreover, DR spectroscopy, unlike spontaneous Raman, is not strongly impacted by fluorescent emission and is thus useful for label-free spectroscopy in environments with high levels of autofluorescence, such as plants. In this Article, we have shown the ability to record Raman spectra from a range of molecular components, some of which are involved in plant metabolic activity, and also demonstrated the ability to differentiate the redox state of NADH/NAD$^+$ with DR spectra. In future work, we will apply DR to imaging molecular compounds with extremely high sensitivity.  Please refer to the Supplementary Information for further experimental details and supporting content.

\section*{Funding Information}
We gratefully acknowledge funding from the W.M Keck foundation and from DOE grants DE-SC0013265 and DE-SC0019545.

\medskip

\noindent\textbf{Disclosures.} The authors RAB and DGW have patented Doppler Raman technology \cite{DRPatent}.

\bibliographystyle{unsrt}  
\bibliography{RFDRPaperBibCarve}

\end{document}


\maketitle
\begin{abstract}
This document provides supplementary information to “Ultrasensitive Doppler Raman spectroscopy using radio frequency phase shift detection”.  The experimental setup for DR spectrscopy is described in detail.  The methods for validating the RF harmonic scaling and GDD scaling relationships of DR are discussed.  A number of different dispersive systems were considered and analyzed for DR.  The measured group velocity dispersion (GVD) of the dispersive systems is presented.  The experimental details for measuring the DR timing jitter power spectral density are discussed and the results are analyzed.  Finally, the exceptional dynamic range of DR is demonstrated and the conversion of an optical frequency shift to a measured voltage is illustrated.
\end{abstract}

\section{Methods}
\begin{figure}[ht]
	\begin{center}
		\resizebox{0.65\linewidth}{!}{\includegraphics[width=0.55\textwidth]{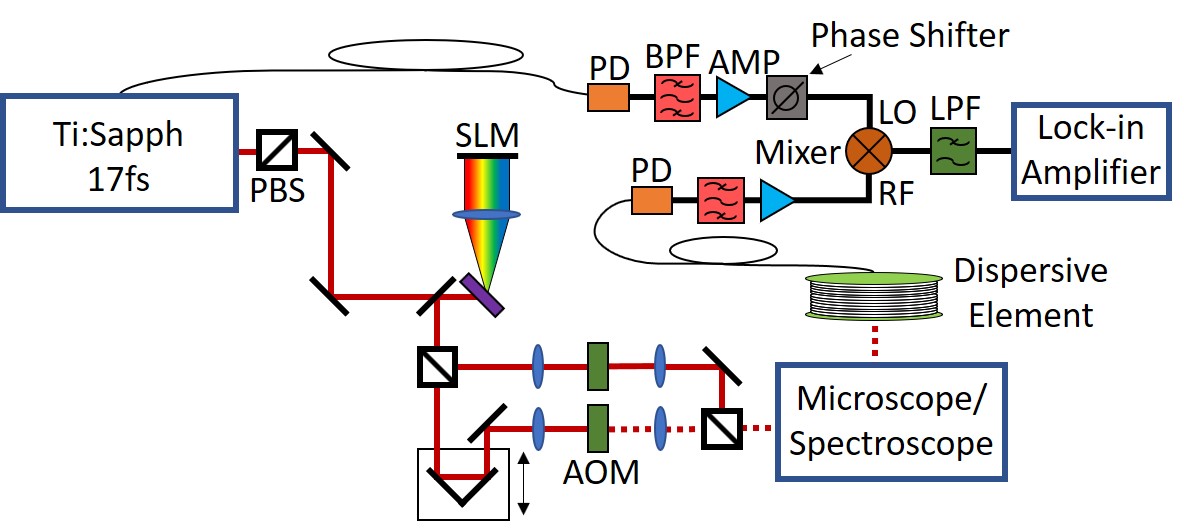}}
		\caption{\label{fig:expSetup}  Experimental setup for DR spectroscopy.}  
	\end{center}
\end{figure}

A schematic of the RF detection DR experimental system is shown in  Fig. (\ref{fig:expSetup}). Here, pulses from a mode-locked Ti:sapphire laser (KM Labs) are sent through an ultrafast pulse shaper to ensure the use of transform-limited pulses in the experiment. The pulses exiting the shaper are directed into a polarization-based Mach-Zehnder interferometer to produce pump and probe pulses with a controllable relative time delay. The intensity of the pump pulse train is modulated with an acousto-optic modulator (AOM) at the modulation frequency $f_\mathrm{mod}$ and the relative pump-probe delay $\tau$ is adjusted with a scanning mirror delay line. The pump and probe pulses are recombined with orthogonal linear polarizations using a polarizing beam splitter. The pump-probe pulse pair is focused into a cuvette with an aspheric lens. The pump and probe pulses are compressed to the transform limit by optimizing the intensity of two-photon absorption fluorescent emission from a bath of yellow highlighter ink diluted in water and placed in the cuvette.  The pulses are recollimated with a matched aspheric lens, then passed through a polarizer to reject the pump pulses. The probe pulse train is coupled into a gradient index multimode fiber with fiber length $L_{\rm fiber}$ and the pulse train power detected with a fiber-coupled photodiode on the distal end of the fiber. Phase sensitive RF detection is accomplished with the $m^{\mathrm{th}}$ harmonic of the oscillator repetition rate ($f_R$) by passing the electronic signal generated by the photodiode through an RF bandpass filter centered at a frequency $\nu_m =  m \, f_R$. 

The filtered RF harmonic is passed through an adjustable RF attenuator before being amplified with a low noise amplifier (LNA), then connected to the RF port of a double balanced RF mixer. A local RF oscillator is derived from a portion of the laser power from the laser oscillator that is filtered and amplified in a manner similar to the RF signal and is injected into the LO port of the mixer. An RF phase shifter is used to adjust the RF and LO signals to be in quadrature, and the IF output of the mixer is low pass filtered, preamplified, and connected to a lock-in amplifier.

As discussed in the main text, the DR spectroscopy Raman spectra in Fig. (4) of the main text agree well with previously reported literature values for biological molecules. \cite{Nishimura1978,Rimai1969,Chen1989,Yue1986,DeGelder2007,Fleming2009,Pawlukojc2005}

\section{Validation of harmonic order and GDD scaling}

As discussed in the main text, the detected RF phase shift in DR spectroscopy is given by $\delta \phi_{\rm RF} = 2 \pi f_r m \varphi_2 \delta \omega$, which indicates that the detected RF phase shift is expected to scale linearly with harmonic order $m$ as well as the amount of GDD ($\varphi_2$) applied by the dispersive element on the probe pulse train detection.

To validate these scaling relationships we performed a series of experiments with bismuth germanium oxide (BGO) as the Raman active sample.  A sequence of MMF fiber spools were used that could be coupled together to vary the GDD applied to the probe arm. For a given harmonic order, $m$, high-quality RF band pass filters (BPF) were used to extinguish adjacent, unwanted, harmonics by at least 50~dB to ensure that a pure sine wave signal was isolated for each harmonic order. At each $m$, the GDD was varied with a combination of fiber lengths, and the amplitude of the detected BGO 90 cm$^{-1}$ wavenumber mode was recorded. The time resolved Raman response was processed using multitaper power spectral density estimation to obtain the the 90~cm$^{-1}$ mode amplitude of BGO.  

\begin{figure}[ht]
	\begin{center}
		\resizebox{0.55\linewidth}{!}{\includegraphics[width=0.45\textwidth]{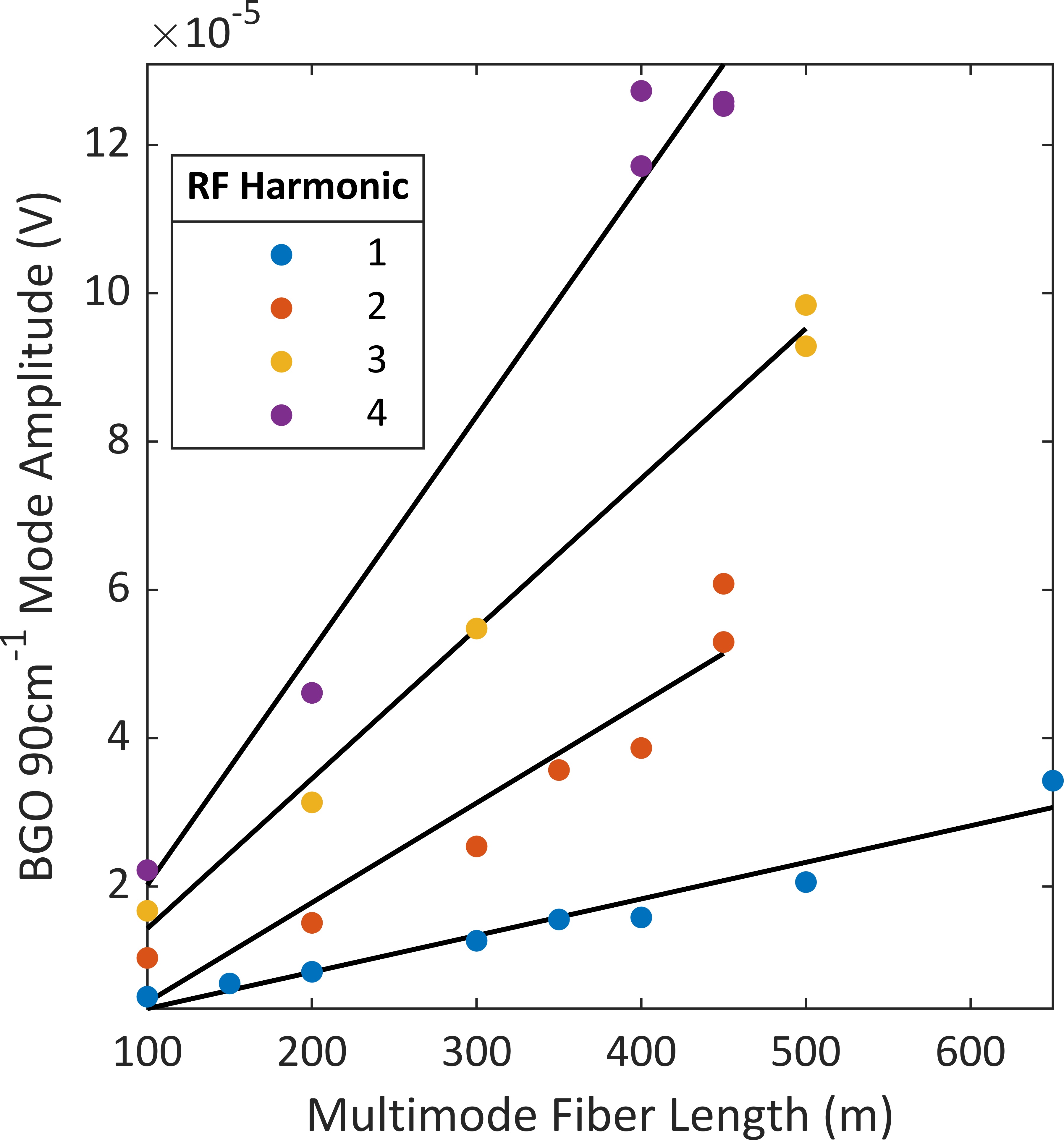}}
		\caption{\label{fig:scaling}  The DR signal scales with RF harmonic order as well as GDD (length of MM Fiber)}  
	\end{center}
\end{figure}

Consistent probe and pump powers were carefully maintained while varying the fiber length for a given harmonic $m$.  Because additional fiber length further chirps out the probe pulses, i.e., stretches the temporal pulse duration, the RF power generated (with a constant probe optical power) in a given harmonic drops with increased fiber length (increased $\varphi_2$).  A single frequency mixer, operating in saturation, was used as a phase detector across the frequency range encompassing the first four RF harmonics of our laser oscillator. To ensure consistent operation from the phase detector (consistent conversion of phase to voltage) the changes in generated RF power with varying fiber lengths and RF harmonic need to be compensated to maintain consistent RF injection power into the mixer. Consistent RF powers injected into the mixer ports are ensured by  a variable RF attenuator that is placed in series with the low noise amplifiers used to amplify the RF signals to the saturation power levels required by the mixer. In this manner, measurements were performed with a series of RF harmonics and fiber lengths that could be quantitatively compared. The recovered amplitude of the 90~cm$^{-1}$ mode of BGO is plotted as a function of the fiber length (GDD) in Fig. (\ref{fig:scaling}) for the first four RF harmonics.  As expected, the 90~cm$^{-1}$ mode amplitude increases linearly with fiber length for a given harmonic.  Additionally, for a given fiber length, the signal increases linearly with harmonic order.  

\section{Measurement of GVD of several dispersive systems}

In DR spectroscopy, it is advantageous to apply as much GDD as possible in order to impart the largest possible signal amplification.  While exploring various potential dispersive systems we directly measured the group delay for the dispersive systems.  For these measurements, the liquid crystal spatial light modulator in the pulse shaper was removed and replaced with a narrow spatial slit transmission filter. The slit passed a narrow, $\approx2~$nm FWHM, band of the optical power spectrum.  The spectrally filtered light was then coupled through the dispersive system being measured and then was directed onto either a standard photodiode (PD) or an avalanche photodiode (APD).  The signal from the photodetector was digitized on a high-speed (8GHz) oscilloscope that was externally triggered by a separate photodiode illuminated by the reference pulse from the oscillator.  

\begin{figure}[ht]
	\begin{center}
		\resizebox{0.55\linewidth}{!}{\includegraphics[width=0.45\textwidth]{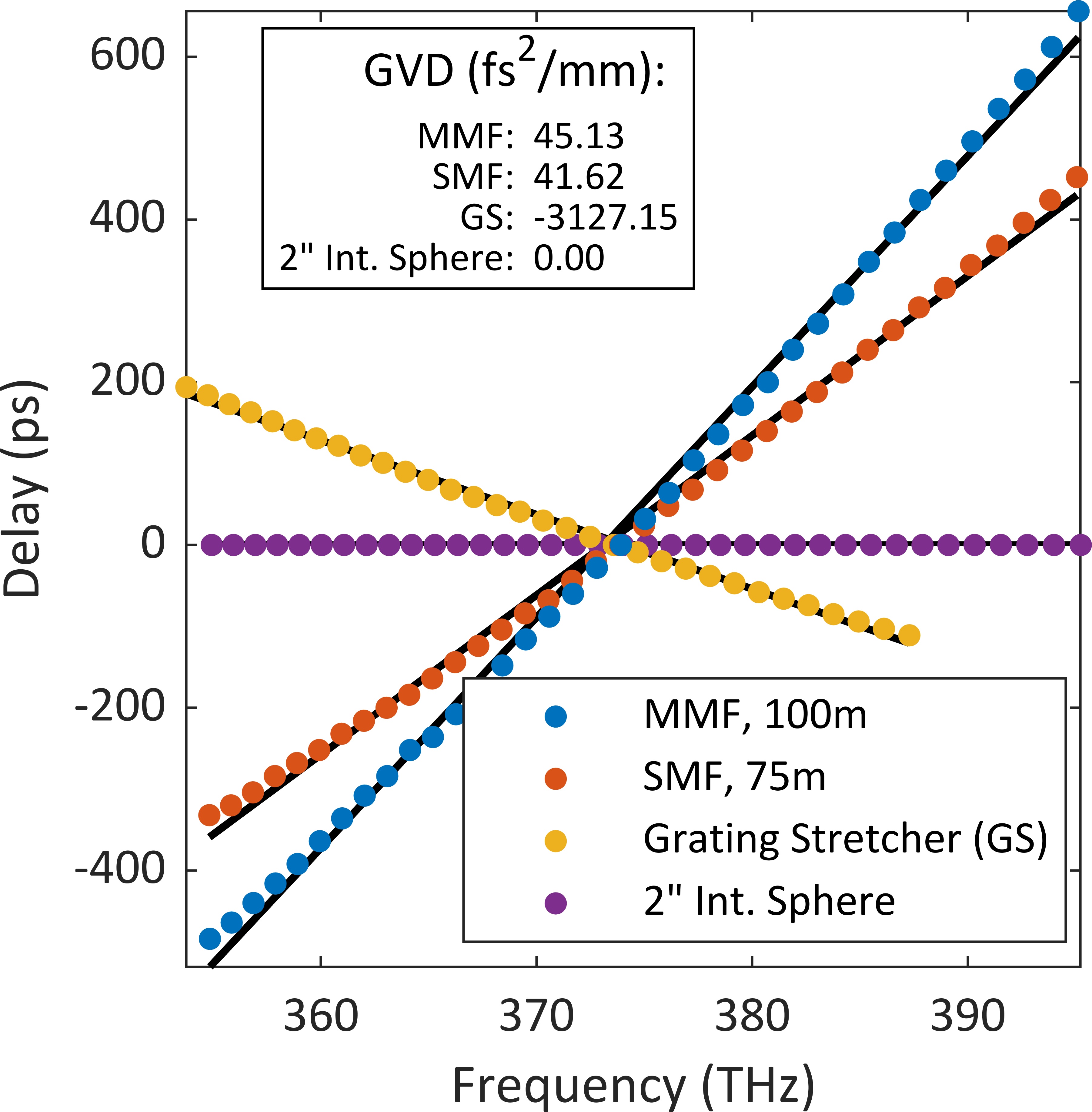}}
		\caption{\label{fig:groupDelay}  Measured group delays of dispersive systems.}
	\end{center}
\end{figure}

Using a linear translation stage, the narrow slit in the conjugate plane of the Martinez compressor was translated across the angularly dispersed bandwidth of our laser oscillator.  For each position of the translation stage, a waveform was recorded on the high-speed scope.  As the slit swept across the spectrum, different narrow-band frequencies of light were passed through the dispersive systems and their relative (to the externally triggered signal) arrival times could be recovered from the recorded waveforms on the scope.  The relative location of the centroid of the waveform is plotted in Fig. (\ref{fig:groupDelay}) as a function of optical frequency.  As can be seen from Fig. (\ref{fig:groupDelay}), the multimode fiber (MMF) had a slightly higher group velocity dispersion (GVD) than the single mode fiber (SMF) that we tested.  In addition to the higher GVD, the MMF is experimentally preferable to work with since it is much easier to couple the post-sample probe light into the MMF fiber. 

Dispersive systems beyond the fiber-based designs were also explored.  We measured the effective GVD for a grating stretcher built with reflective diffraction gratings.  The grating spacing was maximized to 46.5~cm, a limit imposed by the broad 17~fs spectrum and the physical size of the second grating to fully capture the angularly dispersed bandwidth from the first grating.  As can be seen in the Fig. (\ref{fig:groupDelay}), both fiber-based dispersive systems allowed for more GDD than the grating stretcher.  Finally, we explored the possibility of using an integrating sphere as a dispersive system.  We measured pulses exiting the integrating sphere using high-speed 8~GHz PDs (EO Tech) and found that the pulses appeared to be significantly chirped out to several nanoseconds from the pulses entering the integrating sphere which were on the order of femtoseconds.  However, because the distribution of the scattering paths do not impose chromatic dispersion, and there is no resonant scattering in the integrating sphere, the difference in transit times through the integrating sphere across the frequencies within our laser bandwidth displayed no relative difference in arrival time, and thus no change in transit time with pulse center frequency shift. When the pulse enters the integrating sphere and scatters off the inner surface, multiple copies are created that continue to scatter and arrive at the photodiode distributed over several nanoseconds (depending on the radius of the integrating sphere), producing a photodiode waveform with a fast rise time and a long tail with a FWHM of several nanoseconds. With these results, we decided on using lengths of MMF as our dispersive system, a choice that combined the large GDD of the MMF with relative ease of use.

\section{Characterization of timing jitter noise power spectral density}

\begin{figure}[ht]
	\begin{center}
		\resizebox{0.55\linewidth}{!}{\includegraphics[width=0.45\textwidth]{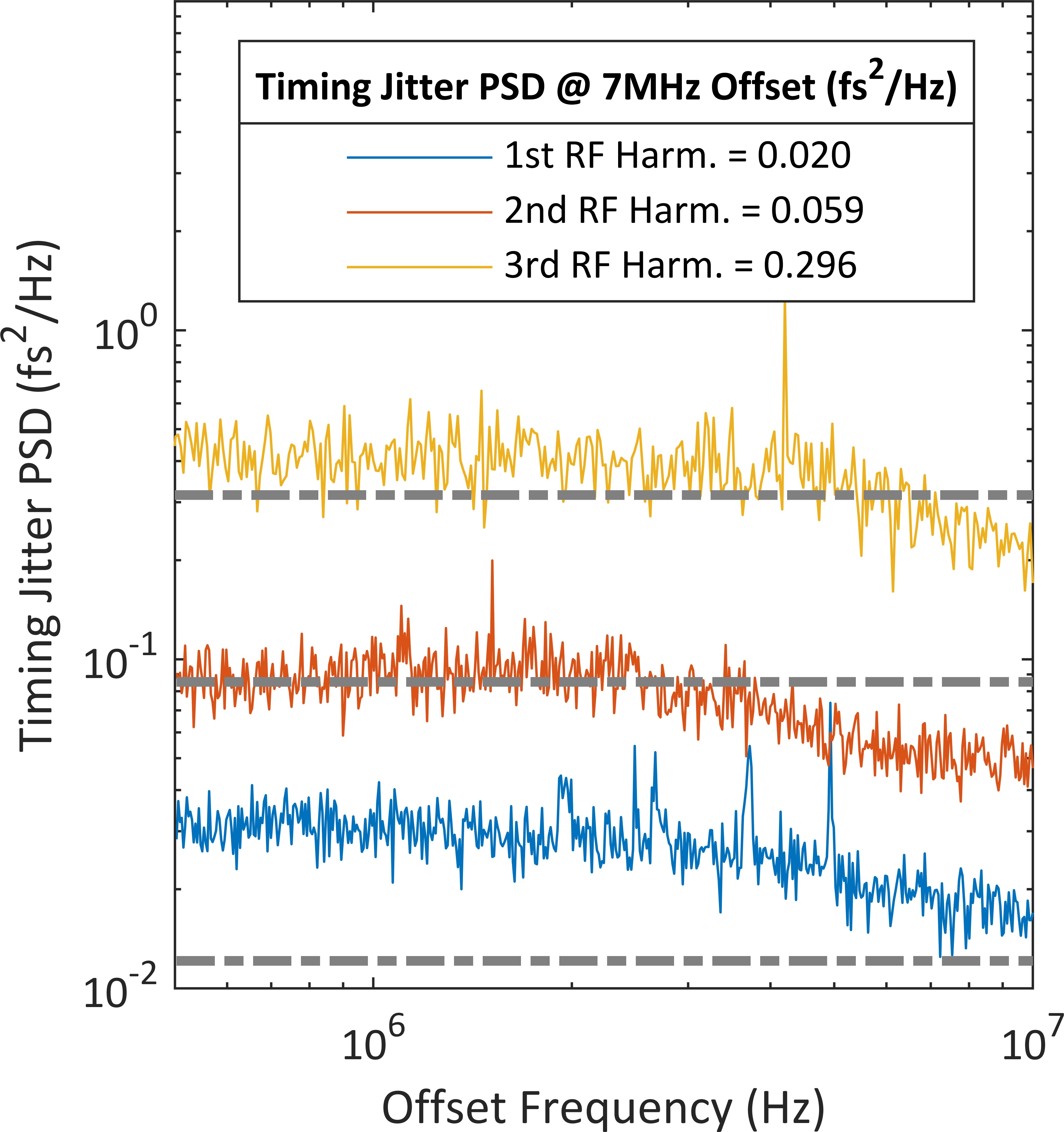}}
		\caption{\label{fig:timeJitterPSD}  Measured timing jitter spectral densities following the procedure outlined in the text. The dashed lines are the phase Shot noise levels computed from the measured RF power in each harmonic order using the formula $S_{\delta \phi} = 10 \log_{10} \left[ h \nu/ 2 \, p_{\rm RF} (\pi \, m\ f_R)^2 \right]$.}
	\end{center}
\end{figure}

Fig. (\ref{fig:timeJitterPSD}) shows timing jitter power spectral densities (PSD) of the DR system for the first three RF harmonics from offset frequencies, $f_{\rm off}$, of 500kHz to 10MHz. The dashed lines represent the shot noise-limited timing jitter PSD expected based on the RF power generated by the PD for each of the harmonics. The timing jitter PSD of the DR system were measured by removing the cuvette from the spectroscopic system, but still coupling the probe light into the MMF fiber as usual.  

The electronic setup was similar to the DR spectroscopy system.  Briefly, a probe pulse was input into a MMF attached to a fiber coupled 1GHz PD and the harmonic of interest is isolated with a high-quality RF BPF.  The filtered signal is then attenuated and/or amplified to be conditioned to the appropriate RF power for the signal/RF port of the frequency mixer being used.  A small portion of the light from the laser oscillator is sent in parallel to an identical fiber coupled PD and is then filtered with an identical high-quality BPF.  This filtered signal is then attenuated and/or amplified to appropriate RF power levels to drive the local oscillator (LO) port of the frequency mixer, which is being run in saturation as a phase detector.  We now have a synthesized RF signal serving as a reference (LO) and the signal generated with probe light creates the signal for the RF port of the frequency mixer.  To ensure the two signals entering the frequency mixer are in quadrature, an RF phase shifter is employed on the reference arm (LO) to tune the quadrature point.  The output of the frequency mixer/phase detector is low-pass filtered, preamplified, and then connected to a spectrum analyzer (Signal Hound).  The noise PSD measured by the spectrum analyzer can be converted to single sideband phase noise using the conversion factor, $K_{\phi}$, of the specific mixer being used.  The conversion factor of the mixers used were measured using the well-known beat-note method.  

Once the single sideband phase noise, $S_{\delta \phi}$, has been measured, it is straightforward to convert this to a timing jitter PSD, which are related by  $S_{\delta\tau_{\rm rms}} = (2 \pi m f_R)^{-2} S_{\delta \phi}$.  As can be seen in Fig. (\ref{fig:timeJitterPSD}), when the DR system is running with the $m=2$ and $m=3$ RF harmonics it is close to the shot noise-limited level.  The separation between the measured noise floor for $m=1$ and the shot noise-limited level can most likely be attributed to amplitude to phase noise conversion in the fiber-coupled PDs.  Our measurements of the timing jitter PSD at the various RF harmonics confirm that we are running DR spectroscopy with the lowest possible timing jitter for phase detection of RF signals synthesized from standard fiber-coupled PDs.  Alternative methods of phase detection promise to reduce the timing jitter PSD noise floor by several orders of magnitude, offering a path to improved detection sensitivity using the principles of DR spectroscopy.  

Under conditions of the lowest noise operation, we produced RF powers of 117, 4.17, and 0.50 $\mu$W for the $m = 1,2,3$ harmonics respectively in our system. For these RF powers, the effective timing jitter PSD is $S_{\delta\tau} = 0.020, 0.058, 0.30 $ fs$^2/$Hz, respectively for increasing harmonic order. In the form $S_{\delta\tau} = 10^{-b}$ fs$^2$/Hz, we find $b = 1.7, 1.2, 0.53$, respectively. The rms timing jitter that sets our detection limit in a $\Delta t$ integration time is $\delta\tau_{\rm rms} = \{0.10, 0.17, 0.39\}/\sqrt{\Delta t}$ fs s$^{-1/2}$, for the three ascending harmonics.

The second and third order harmonics both exhibited a noise floor within $< 0.5$ dB of the the Shot noise limit, whereas the first order harmonic was 2.2 dB higher than the Shot noise limit. Detection with lower harmonic orders produces a higher SNR because the RF power drops nonlinearly with harmonic order due to the highly chirped pulses, which leads to a commensurate nonlinear increase in the noise floor.  The current experiment is limited in sensitivity by the limited RF power generated in the probe pulse photodiode. This RF power can be scaled to produce a lower detection timing jitter PSD as generated in experiments carried out at NIST, where a high power diode detected microwave tones from mode-locked oscillators with a phase noise down to -179 dBc/Hz \cite{Quinlan:2013df}.

\section{Measurement of the noise floor and dynamic range with deionized (DI) water}

\begin{figure}[htbp]
	\begin{center}
		\resizebox{0.55\linewidth}{!}{\includegraphics[width=0.45\textwidth]{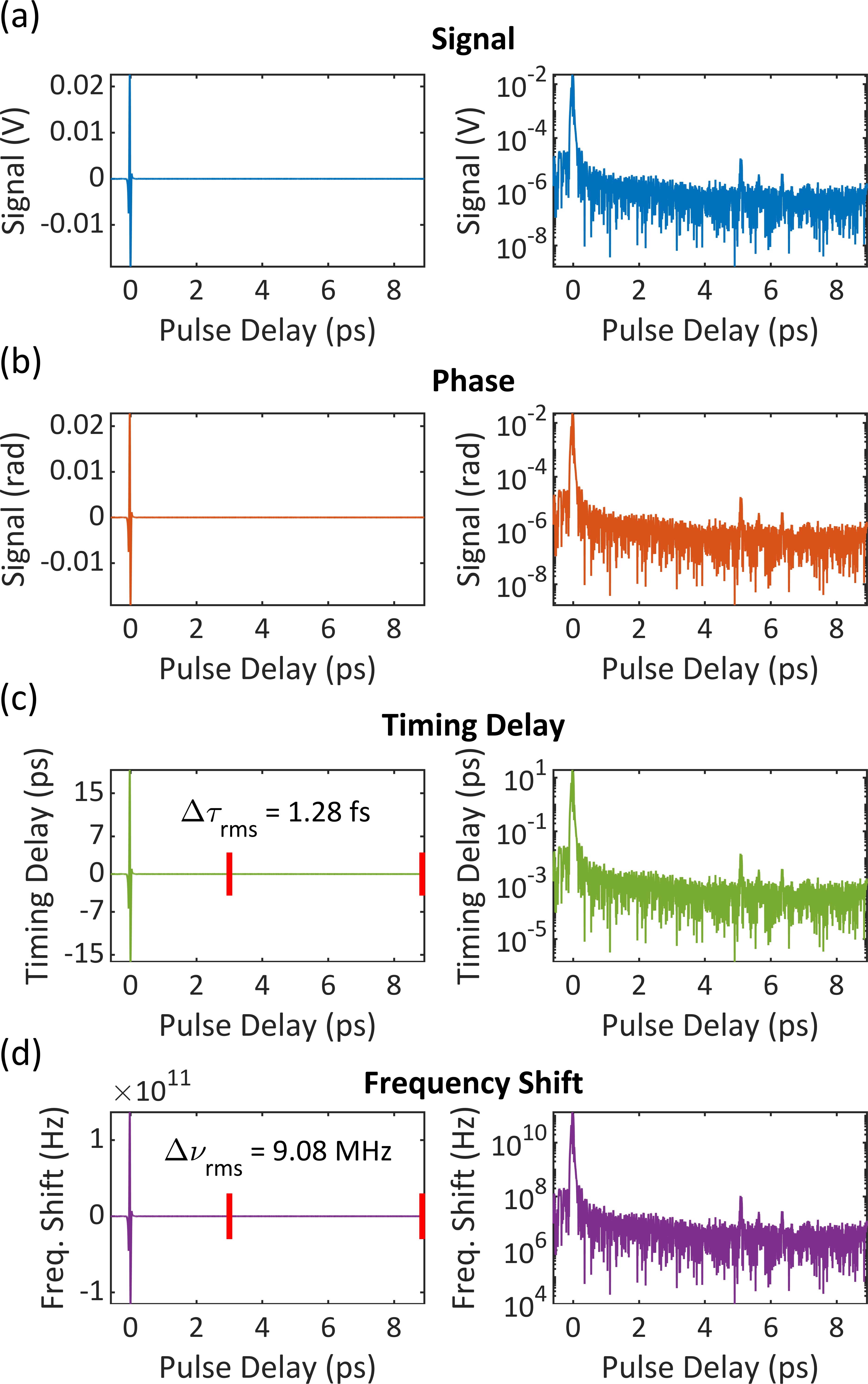}}
		\caption{\label{fig:sigToFreq}  DR system dynamic range and noise floor, with a linear scale in the left column and a log scale in the right column. (a) The measured voltage signal out of the lock-in amplifier, divided by the voltage gain of the pre-amplifier that increased the signal exiting the electronic phase detector. (b) The phase signal obtained from the lock-in voltage level divided by $K_{\phi}$. (c) The timing delay change obtained by dividing the phase noise by the RF harmonic frequency. (d) The extracted frequency shift of the laser pulse obtained by dividing the time delay by the measured GDD of the MMF disperser.}
	\end{center}
\end{figure}

DR spectroscopy is a measurement technique with a substantial dynamic range. To date, we have recorded signals spanning more than 5 orders of magnitude in a single measurement.  To illustrate this capability, Fig. (\ref{fig:sigToFreq} a) shows the time resolved signal recorded by our custom Matlab script for DI water, a sample without a detectable Raman response with DR spectroscopy.  While the Raman-active vibrations are well beyond the Raman spectral coverage from our impulse excitation, there is still a significant phase modulation near $\tau = 0$ from the cross phase modulation between the pump and probe pulse.  This cross phase modulation signal is visible in the region near $\tau = 0$ in Fig. (\ref{fig:sigToFreq} a) and the system quickly relaxes to the DR system noise floor as the delay between the pump and probe pulse is increased. 

Fig. (\ref{fig:sigToFreq} a) shows a typical time-resolved signal recorded in volts by the lock-in amplifier.  The left plot is on a linear y-scale while the right plot is on a log y-scale to highlight the significant dynamic range.  To demonstrate the signal conversion that occurs in DR spectroscopy, Fig. (\ref{fig:sigToFreq} a - d) shows the steps of how one obtains a voltage signal from a frequency shift in the sample.  The two plots obtained in Fig. (\ref{fig:sigToFreq} b) are obtained by simply using the mixer phase detector constant $K_{\phi}$ to convert from volts to radians.  The $K_{\phi}$ for this mixer and its chosen operating conditions is $K_{\phi} = 0.99~\mathrm{V/rad}$. Fig. (\ref{fig:sigToFreq} c) shows the timing delay obtained from the RF phase shift using relation  $\Delta\tau = \phi_{RF}/(2 \pi m f_R)$ where $f_R= 94~\mathrm{MHz}$ and $m = 2$.  $\Delta\tau_{rms} = 1.28~\mathrm{fs}$ is the recorded RMS timing delay over the region representing the noise floor of the system (well removed from the cross phase signal near $\tau = 0$) which is marked by the two red hash marks. Fig. (\ref{fig:sigToFreq} d) shows the frequency shift obtained by converting from timing delay to frequency shift using the GDD of the dispersive system, i.e., the MMF optical fiber.  In this experiment, the GDD of the MMF was measured to be $22.5~\mathrm{ps}^{2}$, thus the absolute frequency shift can be expressed as $\delta \nu = \Delta \tau / (2 \, \pi \, \varphi_2)$.  The RMS frequency shift well-removed from the cross phase signal is shown to be $\Delta\nu_{rms} = 9.08~\mathrm{MHz}$, where the region used to calculate the RMS frequency shift is again indicated by the red hash marks.  Because the region between the red hashes represents the DR spectroscopy noise floor, this RMS frequency shift represents the current minimum frequency shift detectable by PD-detection-based DR spectroscopy.  Working backwards from Fig. (\ref{fig:sigToFreq} d) to (\ref{fig:sigToFreq} a) simply illustrates how a frequency shift experienced by the probe pulse spectrum caused by the excited vibrational coherence in the sample is converted to a timing delay using a dispersive element (GDD), then to a phase shift at a given RF harmonic and is finally converted to a voltage detected by the lock-in amplifier using a frequency mixer operating as a phase detector.

\bibliographystyle{unsrt}  
\bibliography{RFDRPaperBibCarve}
